\documentclass[prd,nofootinbib]{revtex4-1}
\usepackage{amsmath,amssymb}
\usepackage{slashed,verbatim,graphicx}
\usepackage{color}

\newcommand\ket[1]{| #1\rangle}
\newcommand\bra[1]{\langle #1|}
\newcommand\braket[2]{\langle #1|#2\rangle}

\newcommand{\beq}{\begin{equation}}
\newcommand{\beqs}{\begin{equation*}}
\newcommand{\eeq}{\end{equation}}
\newcommand{\eeqs}{\end{equation*}}
\newcommand{\CO}{  {\cal O}  }

\begin{document}
\setlength{\unitlength}{1mm}
\title{Localized states in global AdS}

\author{David Berenstein$^\dagger$, Joan Sim\'on $^\ddagger$}
\affiliation { $^\dagger$ Department of Physics, University of California at Santa Barbara, CA 93106\\
$^\ddagger$ School of Mathematics and Maxwell Institute for Mathematical Sciences,\\
	University of Edinburgh, Edinburgh EH9 3FD, UK}

\begin{abstract} 
We construct both local states and scattering states with finite energy in global AdS by inserting properly regularized operators in the CFT of arbitrary conformal dimension $(\Delta)$ at an instant of time. We give the state fixed angular momentum $(\ell)$ by integrating the result over a sphere with appropriate spherical harmonics. The energy of the states and their angular resolution is computed with CFT operator methods and is independent of having an AdS interpretation.
In the semiclassical limit of large conformal dimension operators, these correspond to single particles localized within
subAdS scales with width $1/\sqrt{\Delta}$ in AdS units, whose subsequent evolution is controlled by bulk geodesics. Our construction allows us to place a particle in any desired geodesic. For radial geodesics, we show that the amplitude to produce the desired state can be thought of as a regularized tunneling amplitude from the boundary to the radial turning point of the radial geodesic, while for other geodesics we argue that the insertion is  at the outermost radial turning point of the corresponding geodesic. 

\end{abstract}

\maketitle

\section{Introduction }
\label{S:Introduction}

In the AdS/CFT correspondence dictionary, one is supposed to excite fields in the bulk by inserting local operators on the boundary 
\cite{Witten:1998qj}. 
The signals then travel from the boundary to the interior where interactions might happen some time later.
For most of these operators however, the naive insertion 
of the operator is meaningless: they produce a non-normalizable state in the bulk and therefore inject an infinite 
amount of energy into AdS (see for example \cite{Berenstein:2014cia}). 
Moreover, these operators are usually irrelevant operators so that even renormalization in the CFT 
becomes problematic. These operators need to be regularized in some form to render the answers meaningful.

At least perturbatively, some of these issues are resolved by the HKLL construction \cite{Hamilton:2005ju,Hamilton:2006az}. This is a prescription where, essentially, one integrates over time in the boundary with some profile and arranges for the high frequency mode amplitudes to interfere between different times so that after a quench protocol they are not excited. It needs to be corrected order by order. The protocol, at least naively, has an explicit dependence on the background if one goes from say,  vacuum AdS, to a 
black hole.  It also has the disadvantage of not being local in time. In this sense, the construction complicates the naive intuition that one would like to have, for example, when preparing a scattering experiment of widely separated particles in the bulk and letting them fall to a common meeting place where the experiment would take place. This is a thought experiment that is used to test holography and locality in the bulk \cite{Gary:2009ae,Gary:2009mi,Heemskerk:2009pn}. In many examples, this locality takes place in higher dimensions. Being able to probe physics of these extra dimensions is interesting, but large momentum in the extra dimensions translates to having dual operators with large conformal dimension $\Delta$.
To address these issues requires understanding the problem of non-normalizability of states in the CFT and the potential infinite injection of energy that might lead to unphysical results.

Large conformal dimension operators are also interesting for other reasons. They can be argued to be dual to particles in the bulk whose
Compton wavelength is much smaller than the AdS radius. As such, one can imagine studying local physics with particles that are 
separated from each other by distance scales that are much smaller than the AdS radius, but still very long compared 
to their Compton wavelengths. Under such conditions a classical or semiclassical analysis could 
give a complete description of the physics for various set-ups, especially if the goal is to directly control the impact parameter
of the collision to study physics.

The Mellin representation of four point functions in conformal field theory has been argued to be a route to address these 
issues \cite{Penedones:2010ue,Fitzpatrick:2011ia}, in particular, the relation to Eikonal physics in AdS \cite{Cornalba:2006xm,Cornalba:2007zb}. Eikonal physics in flat space makes use of notions of impact parameter as related to different partial waves in scattering problems.

There is an alternative way to regulate the problem that has been described in papers by Takayanagi et al. \cite{Goto:2017olq,Goto:2016wme,Nozaki:2013wia}. One replaces the operator insertion $\CO$, by a regulated version that has been evolved in Euclidean time an amount $\epsilon$
\begin{equation}
  \CO_{\epsilon}(t)=\exp(-\epsilon H)\, \CO(t)\exp(\epsilon H)\,.
\end{equation}  
Since the regulated operator is defined via an Euclidean evolution operator, it is not strictly local any longer. It should be thought to have some support on an $\epsilon$ dependent spatial section that goes to zero when $\epsilon\to 0$. 

The purpose of this paper is to explore the physics of such insertions further, focusing on global AdS which is more suitable both for scattering experiments and to study finite mass black hole dynamics. 

The organisation of this paper is as follows. In section \ref{sec:operators}, we introduce regularized CFT scalar primary operators whose insertion acting on the vacuum generates the primary and its descendants, with no angular momentum. We identify the descendants maximising these amplitudes and study their gaussian approximations in the limit of large conformal dimension. In section \ref{sec:bulk}, we show these insertions can be interpreted as bulk semiclassical particles sitting at rest at the turning point of a radial geodesic. This provides a geometric interpretation of the CFT regulator $\epsilon$ and maps the width of the CFT gaussian amplitudes to bulk localisation properties of these excitations. Considering regularised operator insertions not integrated over the CFT sphere, we extend the bulk localisation properties to the angular directions in section \ref{sec:angular}. In section \ref{sec:momentum}, we extend our analysis to descendants with non-zero angular momentum. We show the same bulk interpretation and localisation properties hold for these states. We finish with a brief summary and discussion of our results in section \ref{sec:conc}.

\section{Regularized operators}
\label{sec:operators}

Given a d-dimensional CFT on a cylinder, consider the insertion of an scalar primary operator at fixed time $t$ on the CFT vacuum integrated over the sphere
\begin{equation}
  \int d\Omega\, \CO_{\Delta}(\theta, t)\ket 0
\label{eq:s-wave}
\end{equation}
As shown in \cite{Berenstein:2014cia}, this insertion generates a set of states that includes the primary $\ket{\CO_\Delta}$ and its descendants $\ket{(\partial_\mu\partial^\mu)^k \CO_\Delta}$ which we call $\ket{\Delta+2k}$ under the operator-state correspondence. The amplitudes to create each of these states are given by
\begin{equation}
  A_{\Delta+2k}\simeq \bra{\Delta+2k}\int d\Omega\, \CO_{\Delta}(\theta, t)\ket 0  
\label{eq:cftamplitude}
\end{equation}
and were shown to be proportional to 
\begin{equation}
  |A_{\Delta+2k}|^2 \propto \frac{\Gamma[k+\Delta] \Gamma[\Delta -\frac d2 +k+1]}{k!\,(\Gamma[1+\Delta -\frac d2 ])^2 \Gamma[k+\frac d2]}\,.
\label{eq:ampsq}
\end{equation}

For large $k$, Stirling's approximation $\Gamma[s] \simeq \exp(s \log(s) -s)$ allows to estimate the amplitudes \eqref{eq:ampsq}
\begin{equation}
  |A_{\Delta+2k}|^2 \propto \exp( (2 \Delta -d) \log(k)+\mathcal{O}(1))  
\label{eq:l0}
\end{equation}
Hence, the individual amplitudes are finite, but the state \eqref{eq:s-wave} will only be normalizable if the sum of the amplitudes squared is convergent. Otherwise, the vector \eqref{eq:s-wave}  is outside the allowed Hilbert space of states of the field theory. The normalizability condition 
\begin{equation}
\sum_{k=1}^\infty k^{2\Delta -d}  <\infty 
\end{equation}
requires the terms in the series to decay faster than $1/k$, implying $2 \Delta <d-1$. Since unitarity in the CFT requires $\Delta\geq \frac {d}{2}-1$, the window of normalizability is small. 

Since the amplitudes \eqref{eq:ampsq} to generate higher states grow polynomially, any small exponential correction will tame the divergence caused by the UV modes and render the norm of the state finite. This suggests to instead consider the operator
\begin{equation}
 \int d\Omega\, \CO_{\Delta, \epsilon}(\theta, t)\ket 0
\label{eq:int-op}
\end{equation}
whose amplitude satisfies 
\begin{eqnarray}
A_{\Delta+2k, \epsilon}&\simeq& \exp(-(2k+\Delta) \epsilon) \bra{\Delta+2k}\int d\Omega\, \CO_{\Delta}(\theta, t)\ket 0 \nonumber\\
 &=& A_{\Delta+2k}\exp(-\epsilon(\Delta+2k))\,.
\end{eqnarray}
It is easy to show that such an operator is given by
\begin{equation}
  \CO_{\Delta,\epsilon}(\theta,t)= \exp(-\epsilon H )\, \CO_{\Delta}(\theta,t)\exp(+\epsilon H )
\label{eq:euc-time}  
\end{equation}
where $H$ is the Hamiltonian of the field theory (which in the state $\ket{\Delta +2k}$ has energy $\Delta+2k$ ). Thus, the cut-off $\epsilon$ responsible for taming the UV divergence in the amplitude can be reinterpreted as evolving the insertion of the operator in the field theory in imaginary time an amount $\epsilon$. This regularisation procedure has been considered in \cite{Goto:2017olq,Goto:2016wme,Nozaki:2013wia} for single trace operators.

If one is interested in understanding the physics of these states in terms of some semiclassical bulk description, it is natural to ask, for a given $\epsilon$, what the most populated state is.
Using Stirling's approximation to evaluate the regularized amplitude $A_{\Delta+2k, \epsilon}$ 
\begin{equation}
  |A_{2k+\Delta,\epsilon}|^2 \simeq \exp\left[({2\Delta- d}) \log (k) -2 \epsilon (2k +\Delta)\right]
\label{eq:large-k-amp}
\end{equation}
its maximum can be determined as the saddle point in $k$
\begin{equation}
  k_{\max} = \frac{2\Delta - d }{4\epsilon}\,.
\label{eq:kmax}
\end{equation}
Since we neglected terms of order $\Delta/k$ and $d/k$ when evaluating \eqref{eq:large-k-amp}, consistency requires $\epsilon\ll 1$. To study the width of the amplitudes $|A_{\Delta+2k, \epsilon}|^2$ around this maximum, the second derivative yields
\begin{equation}
  \partial^2_k \log(|A_{\Delta + 2k,\epsilon}|^2 )= -\frac{2\Delta -d}{k^2}\simeq -\frac{1}{\sigma^2}
\end{equation}
where $\sigma$ is the standard deviation of the approximate Gaussian distribution describing $|A_{\Delta+2k, \epsilon}|^2$ in this regime. One infers the relative uncertainty in $k$ scales as
\begin{equation}
\frac{|\delta k|^2}{k_{\max}^2} \simeq \frac1 {2\Delta -d}
\end{equation}
This is small for operators of large dimension $(\Delta\gg 1)$. Hence, in the regime of $\Delta\gg 1$ and $\epsilon\ll 1$, most of the probability is carried by states with $k_{\max}$ with a narrow gaussian distribution around it.

If instead of parameterising the amplitudes of these states using $k$, we use their energy, this is approximately given by
\begin{equation}
  E\simeq 2 k_{max} +\Delta = \frac{2 \Delta -d} {2\epsilon} +\Delta \simeq \frac{\Delta-d/2}{\epsilon}\,,
\label{eq:q-energy}
\end{equation}
where we used $\epsilon\ll 1$ in the last step, and the relative uncertainly in $E$ scales like the one in $k$
\begin{equation}
  \frac{|\delta E|^2}{E^2}\simeq \frac1 {2\Delta -d}\,.
\label{eq:en-fluc}
\end{equation}

The upshot is that the set of regularized operators \eqref{eq:int-op} produces states of finite energy $E\simeq \Delta/\epsilon$ and small fluctuations when $\Delta\gg 1$ and $\epsilon\ll 1$. These operators can be thought of as producing an initial condition for a state at fixed energy, like scattering states. We will develop this intuition further when we discuss its semiclassical dual AdS interpretation.

\section{The AdS dual interpretation}
\label{sec:bulk}

Our analysis so far generates s-wave states due to the integration over the CFT (d-1)-sphere in \eqref{eq:int-op}, but it otherwise applies to any scalar primary operator of conformal dimension $\Delta$. When the latter is single trace, the standard AdS/CFT dictionary maps it to an scalar bulk field of mass $m$ \cite{Witten:1998qj}
\begin{equation}
  m= \sqrt{\Delta(\Delta -d)}\simeq \Delta -d/2 +\mathcal{O}(1/\Delta)
\label{eq:mass}
\end{equation}
in units of the AdS radius. The excitations generated by this bulk field are expected to be described by an AdS point particle of the same mass $m$ in some appropriate semiclassical regime. What we discuss in this section is the validity of this expectation, allowing us to provide a geometrical interpretation for the CFT cut-off $\epsilon$ in the process\footnote{Our formulae below can be applied to small objects like D-branes wrapped on extra dimensions, but acting like point particles in AdS, and to bound states of many single particles that can effectively behave like a heavier effective particle in AdS.}.

Inserting the operator \eqref{eq:int-op} at $t=0$ should provide an initial condition for a bulk particle in global AdS, the latter describing the dual of the CFT vacuum $|0\rangle$. This particle should initially be at rest. This is because all CFT amplitudes are simultaneously real. Hence, the CFT state is consistent with an initial configuration having time reversal symmetry. The latter can only happen classically for a point particle if the initial velocity vanishes. Due to the s-wave nature of the CFT state, the bulk particle must follow a radial infalling geodesic as time evolves. 

Hence, the bulk description involves
\begin{equation}
  S = m\,\int d\tau\,\sqrt{\cosh^2\rho - \dot{\rho}^2}\,,
\end{equation}
where we already used the global AdS parameterisation
\begin{equation}
  ds^2 = -\cosh^2 \rho \,dt^2 + d\rho^2 +\sinh^2 \rho \,d \Omega_{d-1}^2
\label{eq:global}
\end{equation}
and chosen the gauge $\dot{\tau}=1$. The energy of the bulk particle is conserved
\begin{equation}
  E = m\,\frac{\cosh^2\rho}{\sqrt{\cosh^2\rho - \dot{\rho}^2}}\,.
\end{equation}
Evaluating it at the turning point $\rho=\rho_\star$, where $\dot{\rho}=0$, 
\begin{equation}
  E = m\,\cosh\rho_\star\,,
\label{eq:p-energy}
\end{equation}
and matching to the CFT energy \eqref{eq:q-energy}, one infers
\begin{eqnarray}
  E=m\,\cosh\rho_\star &=& \frac{\Delta - d/2}{\epsilon} + \Delta- d/2 + d/2  
\label{eq:energy} \\
 &  \Rightarrow & \quad \cosh\rho_\star = \frac 1 \epsilon+1\,,
\label{eq:geom}
\end{eqnarray} 
where we already used \eqref{eq:mass}. The expression for $\rho_\star$ gives a physical solution whenever $\epsilon>0$. When $\epsilon$ is not necessarily small, the terms not scaling as $1/\epsilon$ must be kept to provide a more accurate description of the physics. Indeed, once the particle is inserted in some location, it will follow a full radial geodesic associated to that initial condition. The three terms in the energy \eqref{eq:energy} correspond to the kinetic energy of the particle when it crosses the origin (the $1/\epsilon$ term), the rest energy of the particle at the origin (the mass $m\simeq \Delta - d/2$) and the zero point energy for a point particle at the center of global AdS (the d/2 term) \cite{Berenstein:2002ke}. This last term is quantum and should be ignored in a leading semiclassical treatment. If $\epsilon \ll1$, all non-scaling terms can be ignored.

Equation \eqref{eq:geom} provides a re-interpretation of the cut-off $\epsilon$ as a radial position of the bulk excitation at the turning point. In the small $\epsilon$ limit, $\rho_\star\sim \log(1/\epsilon)$. This is consistent with the radial direction being related to the scale size in the dual theory \cite{Susskind:1998dq}. Indeed, the parameter $\epsilon$ carries units of time because it multiplies the Hamiltonian in the regulator. Turning on a small $\epsilon$ is equivalent to having a small scale present. The precise matching between scale and radial direction depends on coordinate choices. This is entirely consistent with the results of Takayanagi et al. \cite{Goto:2017olq,Goto:2016wme,Nozaki:2013wia}.

For this classical interpretation to hold, the quantum fluctuations of the CFT state must be well within the bulk particle's Compton wavelength $\lambda_\text{c}$. Using \eqref{eq:geom}, one can translate the fluctuations in the CFT energy \eqref{eq:en-fluc} into fluctuations in the radial position of the particle at the turning point
\begin{equation}
   \delta \rho_\star \sinh \rho_\star = \frac{\delta E}{m} \simeq \frac{1}{m} \frac{E}{\sqrt{m}}
\end{equation}
leading to the conclusion
\begin{equation}
 1 \gg \delta \rho_\star \simeq \frac 1 {\sqrt m} \gg \lambda_\text{c} \simeq \frac{1}{m}\,.
\label{eq:compton}
\end{equation}
Hence, in the limit of $\Delta\gg 1$, the bulk particle is not only well localized within a radial region that is much smaller than the AdS scale, which is set to one in our analysis, but also within a parametrically larger region than $\lambda_\text{c}$, so that it can be considered at rest, i.e. a particle localized within a region of the size of the Compton wavelength would not be at rest due to the uncertainty principle. 

The uncertainty $\delta \rho_\star \simeq 1/\sqrt{m}$ in \eqref{eq:compton} is compatible with the physics explaining the three terms in the energy \eqref{eq:energy}. A non-relativisitc particle at rest at the origin of AdS has a frequency of oscillation that is independent of the mass of the particle. Locally, the physics is that of a harmonic oscillator, 
\begin{equation*}
    H_{\text{non-rel}}=\frac{\hat p^2}{2m}+\frac{m\omega^2}{2}\hat x^2
\end{equation*}
where the frequency of oscillation is $\omega=1$. Applying virial's theorem to the ground state, one infers $m (\delta \hat x)^2 \simeq 1$. This reproduces the scaling 
$\delta \hat x\simeq 1/\sqrt{m}$ in \eqref{eq:compton} and provides further evidence for the idea that the bulk particle is at rest at the point of insertion. After insertion, the particle will follow an infalling radial geodesic closely. Changing the radial localisation properties of the bulk particle as an initial condition would require to change the details of the regulator. This is beyond the scope of the present paper.

To sum up, operators \eqref{eq:int-op} create particles of mass $m \sim \Delta $ at location $\rho_\star$, which are initially at rest. Their CFT time evolution corresponds to the bulk particles falling towards the origin of AdS along radial geodesics. This is how the action of the CFT time evolution gets mapped into classical motion in AdS in this semiclassical approximation.  As such, this construction provides a natural candidate for preparing scattering states of finite energy that are well localized on regions smaller than the AdS radius and focusing towards the origin of AdS.

Additional intuition for this interpretation can be obtained from the CFT probability amplitude itself in \eqref{eq:large-k-amp} evaluated at its maximum \eqref{eq:kmax}. Its dominant contribution 
\begin{equation}
  A_{2k_{max}+\Delta} \simeq \exp(  (\Delta - \frac d2) \log (1/\epsilon)+\mathcal{O}(1) ) 
\label{eq:max-amp}
\end{equation}
can be written in terms of the turning point $\rho_\star$, using \eqref{eq:geom}, as 
\begin{equation}
  A_{2k_{max} +\Delta} \simeq \exp( m \rho_\star) \, .
\end{equation} 
The quantity $\rho_\star$ is the spacelike geodesic length from the turning point to the origin. It can be thought of as the negative of a regularized spatial geodesic length from $\rho_\star$ to the AdS boundary at a fixed time $(\dot{\tau}=0)$ 
\begin{equation}
   \tilde{L}_\star =\lim_{\rho_\infty\to \infty} \left[\int^{\rho_\infty}_{\rho_\star} ds\, \sqrt{\dot{\rho}^2} -\rho_\infty\right] = -\rho_\star\,,
\label{eq:l-star}
\end{equation}
so that
\begin{equation}
  A_{2k_{\text{max}} +\Delta} \simeq \exp(- m \tilde L_\star) \,.
\label{eq:tunn}
\end{equation} 
This observation suggests the amplitude itself may be interpretable in terms of a tunneling computation. We will see this intuition is accurate when $\epsilon$ is small.

\subsection{A tunneling calculation}
\label{sec:tunn}

To check whether the suggested tunneling interpretation is accurate, one must check whether the amplitude \eqref{eq:tunn} agrees with the (properly regularised) semiclassical Euclidean AdS particle action and whether the Euclidean time evolution appearing in the regularisation \eqref{eq:euc-time} matches the bulk euclidean time between the boundary and the turning point $\rho_\star$, which sets the initial condition for the bulk interpretation in section \ref{sec:bulk}.

To prepare the state of a massive particle at rest located at $\rho_\star$ via a path integral, consider the euclidean action principle
\begin{equation}
  S_{\text{ec}}= m \int \sqrt{ \dot \rho^2 + \cosh^2 \rho \,\dot \tau^2}\, ds\,.
\end{equation}
Because of the initial condition, the energy of this particle equals \eqref{eq:p-energy}. Choosing the gauge $\dot \tau=1$, the euclidean time that it takes to get to the boundary $\rho_\infty$ equals
\begin{equation}
  \int^\tau_0 d\tau =  \lim_{\rho_\infty\to \infty} \int^{\rho_\infty}_{\rho_\star} \frac{ \cosh\rho_\star \,d\rho}{\cosh\rho\sqrt{\cosh^2 \rho-\cosh^2 \rho_\star}}\,.
\label{eq:full-time}
\end{equation}
Since our semiclassical considerations many times require $\rho_\star \gg 1$, this euclidean time yields
\begin{equation}
  \tau \simeq 2\,e^{-\rho_\star} \lim_{\rho_\infty\to \infty} \int^{\rho_\infty}_{\rho_\star} \frac{e^{2(\rho_\star-\rho)}\,d\rho}{\sqrt{1-e^{2(\rho_\star-\rho)}}} = 2\,e^{-\rho_\star} \simeq \frac{1}{\cosh\rho_\star} 
  =\epsilon\,.
\label{eq:bulk-time}
\end{equation}
Thus, the second of our requirements for a proper tunneling interpretation is fulfilled.

Let us compute the on-shell Euclidean action for a point particle starting at $\rho_\star$ and reaching the boundary at $\tau$, as in \eqref{eq:bulk-time}. Since this action is proportional to the geodesic length connecting these points, it will diverge\footnote{It is worth stressing that this geodesic is \emph{different} from the one giving rise to \eqref{eq:l-star}, even though it is regulated below using the same $\tau=0$ geodesic.}. To regulate it, we cut it off at $\rho_\infty$, subtract the length of the spacelike geodesic from the AdS origin till $\rho_\infty$ at constant $\tau=0$ and take the limit $\rho_\infty$ afterwards
\begin{equation}
   L_\star =\lim_{\rho_\infty\to \infty} \left[\int^{\rho_\infty}_{\rho_\star} d\rho\frac{  \cosh(\rho) }{\sqrt{ \cosh^2 \rho-\cosh^2\rho_\star}}-\rho_\infty\right]\,.
\end{equation}
Taking the limit yields
\begin{equation}
  L_\star =-\log(\sinh \rho_\star)\simeq -\rho_\star\equiv \tilde L_\star\,.
\end{equation}
Hence, the amplitude \eqref{eq:tunn} equals
\begin{equation}
  A_{2k_{\max} +\Delta} \simeq \exp(- S_{\text{ec}})\,.
\label{eq:tunn}
\end{equation} 
Both checks together provide strong evidence for the tunneling interpretation of the CFT amplitudes \eqref{eq:tunn}. 

Before closing this discussion, notice the exact calculation of the euclidean time integral \eqref{eq:full-time} yields
\begin{equation}
  \sinh \tau = \frac{1}{\sinh\rho_\star}\,.
\label{eq:fulltau}
\end{equation}
This result is required for the evaluation of the two point function, which in the current Euclidean setup must behave like
\begin{equation}
    \langle \CO(\theta,+\tau)\CO(\theta,-\tau)\rangle \sim \left(\sinh\tau\right)^{-2\Delta}\,.
\end{equation}
According to the standard AdS/CFT dictionary, the 2-pt function should equal $\exp(- S_{\text{ec}})$, where the full euclidean action corresponds now to the euclidean geodesic connecting the two boundary points in the correlator. Observing that $S_{\text{ec}} = 2m\,L_\star = 2m \log \sinh \tau$, we recover the required behaviour, after using \eqref{eq:fulltau}, except for the substitution $m\to \Delta\sim m+d/2$. As discussed previously, the $d/2$ correction is a zero point energy \cite{Berenstein:2002ke}, which should arise in the semiclassical computation from a one loop correction to the effective action. 

\subsection{Angular localization}
\label{sec:angular}

Our discussion so far focused on the radial localisation properties of our state \eqref{eq:int-op}. If, instead of integrating over the sphere, we would have considered a regulated local operator insertion 
\begin{equation}
  \ket{\theta,\epsilon} = \exp(-\epsilon H) \CO_{\Delta}(\theta,0) \ket 0 \equiv \CO_{\Delta,\epsilon}(\theta,0) \ket 0\,,\label{eq:reg_local}
\end{equation}
generating an excited state on the vacuum at $t=0$, we could study its localisation properties in the angular direction. Using the CFT 2-pt function on the cylinder, the inner product of the regulated states \eqref{eq:reg_local} is exactly given by (see for example \cite{Berenstein:2014cia})
\begin{equation}
 \braket{\theta_2 ,\epsilon}{\theta_1,\epsilon} =\langle \mathcal{O}(\epsilon, \theta_1)\,\mathcal{O}(-\epsilon, \theta_2)\rangle_{\text{cyl}} = \frac{1}{\left[e^{2\epsilon} + e^{-2\epsilon} - 2\cos\delta\theta\right]^\Delta}\,,
\end{equation}
where $\delta\theta = \theta_1-\theta_2$ is the relative angle between the two insertions. Notice, in particular, that when $\epsilon\sim\delta\theta \ll 1$ we can approximate this as
\begin{equation}
  \braket{\theta_2,\epsilon}{\theta_1,\epsilon} \approx \frac{1}{(2\epsilon)^{2\Delta}\,\left(1+\frac{\delta\theta^2}{4\epsilon^2}\right)^\Delta} \approx
  e^{-2\Delta\log\epsilon}\,e^{-\Delta (\delta\theta)^2/(4\epsilon^2)}\,,
\label{eq:trans_unc}
\end{equation}
where in the last step we considered $\delta\theta \ll \epsilon$. The first exponential matches the tunneling amplitude \eqref{eq:max-amp}, suggesting this state remains localized at the same $\rho_\star$ as in \eqref{eq:geom}. The second exponential describes the localization in the angle $\theta$. Indeed, it is a gaussian of size $\delta\theta \sim \frac{\epsilon}{\sqrt{\Delta}}$, which is indeed much smaller than $\epsilon$ in the large $\Delta$ limit.

These observations suggest that operators $\CO_{\Delta,\epsilon}(\theta,t)$ create localized states in both the radial and the angular directions that are controlled by the cut-off, and that they should 
all have the same energy $E$. To check the angular localization, consider the fluctuation in the proper angular distance $(\delta x_\perp)$ in AdS at the insertion point $(\rho_\star,\theta)$ 
\begin{equation}
    \delta x_\perp = \sinh \rho_\star \delta \theta \simeq \frac 1{\sqrt{\Delta}}
\end{equation}    
after using the small $\epsilon$ approximation in \eqref{eq:geom}. This is of the same order of magnitude as the one computed in \eqref{eq:compton} for $\delta \rho_\star$. Hence, the bulk description can again be interpreted as a particle that is initially at rest, modulo uncertainty principle considerations. To check 
this interpretation of a particle at rest, notice that since the angular distribution is effectively gaussian (see \eqref{eq:trans_unc}), the particle should be considered as a minimal uncertainty packet. It follows the uncertainty in the conjugate variable to $\delta x_\perp$  is 
\begin{equation}
  \delta p_\perp \simeq \sqrt{\Delta}\,.
\end{equation}
This uncertainty corresponds to an uncertainty in the velocity transverse to the radial direction given by
\begin{equation}
    \delta v_{\perp} \simeq \frac 1m \delta p_{\perp} \sim \frac{1}{\sqrt{\Delta}}\,.
\end{equation}
This, together with our previous radial localization discussion, shows the particle is non-relativistic at large mass $m$ in all directions of motion at the insertion point $(\rho_\star,\theta)$.

\section{Adding angular momentum}
\label{sec:momentum}

To describe more general states not being at rest at the bulk insertion point, we can simply consider operators $\CO_{\Delta}(\theta,t)$ acting on the vacuum. These generate all possible descendants of the primary state $\CO_\Delta$, not just those at vanishing angular momentum $\ell=0$. Indeed, if we insert one of these operators in the Euclidean plane, the Taylor series
\begin{equation}
    \CO_\Delta(x)= \sum \frac 1 {[a]!} \partial^{[a]}  \CO_\Delta(0)
\end{equation}
generates a series for all derivatives of $\CO_\Delta$. This is a generating function for all of the descendants of the primary $\CO_\Delta$ under the operator state correspondence, which includes the descendants with non-zero angular momentum.
When mapping to the cylinder for radial quantization, other quantum numbers, like the angular momentum, can be fixed by integrating the insertion of $\CO_\Delta(\theta,t)$ in the cylinder with different non-constant angular profiles on the sphere. We expect this procedure to prepare CFT states in any partial wave starting at $t=0$, with an amount of energy controlled by $\epsilon$ once the operator is properly regularized, whose bulk description is specified by infalling bulk particles starting at rest in the radial direction with some fixed angular momentum in the angular directions. These would then fall to the AdS origin where a putative scattering event could take place. 

To compute the amplitudes of these states with a fixed angular momentum, one proceeds similarly to \eqref{eq:cftamplitude}. One starts with $\CO_{\Delta}(\Omega, t)\ket 0$ but this time one projects the state into a fixed angular momentum by integrating with a non-constant spherical harmonic $Y_{\ell}$ into a different partial wave
\begin{equation}
  A_{\Delta+2k+\ell}\simeq \bra{\Delta+2k+\ell}\int d\Omega\, Y_\ell(\Omega) \CO_{\Delta}(\Omega, t)\ket 0  
\label{eq:amp-s}
\end{equation}
using $Y_{\ell}(\Omega)$, one of the spherical harmonics at momentum $\ell$. To get a semiclassical state, one chooses a spherical harmonic that is a highest weight state (e.g. $Y_{\ell\ell}$ in a two dimensional sphere), which can be thought of as having a classical angular momentum $\ell$ on a single plane. 

The standard AdS/CFT dictionary \cite{Fitzpatrick:2011jn,kaplan} suggests to derive the CFT wave-function \eqref{eq:amp-s} as
\begin{equation}
  A_{\Delta + 2k+\ell} \simeq \lim_{\rho\to \infty} \int d\Omega\, Y_\ell(\Omega)  \,\cosh^{\Delta}\rho\, \langle \Delta + 2k + \ell |\Phi(\rho,t,\Omega)\rangle\,,
\label{eq:dictionary}
\end{equation}
where $\Phi (\rho,t,\Omega)$ is the scalar bulk field dual to the scalar primary CFT operator $\CO_\Delta$. This approach was already followed in appendix B of \cite{Goto:2017olq} and extended to an arbitrary number of dimensions in \cite{terashima}.. We reproduce it here in any number of CFT dimensions. The basic idea is to use the properly normalized classical solution to the wave equation in global AdS \cite{Cotaescu:1999wi,Fitzpatrick:2011jn,kaplan}
\begin{equation}
  \Phi (t,\eta,\Omega) = \frac{1}{{\cal N}_{\Delta,k,\ell}}\,e^{-i\omega t}\,Y_{\ell m}(\Omega)\,\sin^{\ell}\eta\cos^\Delta\eta\,{}_2 F_1(\Delta +k+\ell,-k,\ell +\frac{d}{2};\sin^2 \eta)\,,
\label{eq:bulk-scalar}
\end{equation}
with normalization
\begin{equation}
  {\cal N}_{\Delta,k,\ell} = (-1)^k\,\sqrt{k!\,\frac{(\Gamma[\ell+d/2])^2\Gamma[\Delta+k+1-d/2]}{\Gamma[\ell+k+d/2]\Gamma[\Delta+k+\ell]}}\,.
\end{equation}
Notice the radial coordinate $\rho$ in \eqref{eq:global} is related to $\eta$ above by $\sec^2\eta = \cosh^2\rho$. Plugging \eqref{eq:bulk-scalar} into \eqref{eq:dictionary} (and having normalized the descendant state $|\Delta + 2k+\ell\rangle$), the amplitude reduces to
\begin{equation}
  \left|A_{\Delta + 2k+\ell}\right|^2 \propto \frac{1}{{\cal N}_{\Delta,k,\ell}^2}\,\left({}_2 F_1(\Delta +k+\ell,-k,\ell +\frac{d}{2};1)\right)^2.
\end{equation}
The proportionality comes from the spherical integral over the spherical harmonic and the hypergeometric function is evaluated at $1$ because $\eta\to \pi/2$ when we reach the boundary at $\rho\to \infty$, which is the reason why the divergent $\cosh^{2\Delta}\rho$ prescription gets stripped off by the vanishing $\cos^{2\Delta}\eta$ from \eqref{eq:bulk-scalar}. 

Evaluating the hypergeometric function and using identities of gamma functions, the amplitude yields
\begin{equation}
  \left|A_{\Delta + 2k+\ell}\right|^2 \propto \frac{\Gamma[\Delta+k+\ell]\Gamma[\Delta+k+1-d/2]}{k!\,\Gamma[\ell+k+d/2]\left(\Gamma[\Delta+1-d/2]\right)^2}\,.
\label{eq:l-amp}
\end{equation}
This reproduces \eqref{eq:ampsq} when $\ell=0$ and extends it for non s-wave modes. For large $k$, the amplitudes \eqref{eq:l-amp} still grow polynomially in $k$, like in our $\ell=0$ analysis in \eqref{eq:l0}. Regulating the operators as in \eqref{eq:euc-time}, we obtain the regulated amplitudes
\begin{equation}
   |A_{\Delta + 2k+\ell,\epsilon}|^2 = |A_{\Delta + 2k+\ell}|^2\,\exp(-2\epsilon(\Delta+2k+\ell)) \propto  \frac{\Gamma[\Delta+k+\ell]\Gamma[\Delta+k+1-d/2]}{k!\,\Gamma[\ell+k+d/2]\left(\Gamma[\Delta+1-d/2]\right)^2}\,\exp(-2\epsilon(\Delta+2k+\ell))\,.
\end{equation}
To develop the bulk interpretation of these operators, let us analyse the localization properties of these amplitudes in regimes compatible with a semi-classical description. Using Stirling's approximation, 
\begin{equation}
  \log |A_{\Delta + 2k+\ell,\epsilon}|^2 \simeq  (\Delta-\frac d2) \log(k+\ell)+(\Delta-\frac d2) \log\,k-2\epsilon(\Delta+2k+\ell)\,.
\label{eq:stirling-l}
\end{equation}
Since $\ell$ is {\it fixed}, the amplitude is maximised for $k_{\max}$ satisfying
\begin{equation}
  \frac{1}{k_{\text{max}}+\ell}+\frac{1}{k_{\text{max}}} \sim \frac{4\epsilon}{\Delta -d/2}\,.
\end{equation}
If $\ell$ is small, one recovers the previous answer \eqref{eq:kmax}. If $\ell\gg\Delta/\epsilon$, one can ignore the first term and then
\begin{equation}
  k_{\max} \sim \frac{\Delta}{4 \epsilon}
\end{equation}
is half the value for $\ell=0$. To evaluate the relative uncertainty in $k$, we compute the second derivative on $k_{\max}$
\begin{equation}
  \partial^2_k \log(|A_{2k+\Delta,\epsilon}|^2 )= -\left(\Delta -d/2\right)\left(\frac{1}{k_{\max}^2} + \frac{1}{(k_{\max}+\ell)^2}\right) \simeq -\frac{1}{\sigma^2}\,.
\end{equation}
This also has an approximate Gaussian distribution and leads to a relative uncertainty in $k$ scaling like
\begin{equation}
  \frac{|\delta k|^2}{k_{\max}^2} \sim\frac{1}{\Delta}\,,
\label{eq:new-rel}
\end{equation}
for both regimes of angular momentum, i.e. no matter how $\ell$ scales, the large conformal dimension limit $(\Delta\gg 1)$ still controls the localisation of the amplitude.

Following our analysis in section \ref{sec:bulk}, one expects states $|\Delta + 2k_{\max} + \ell\rangle$ to be captured by classical geodesics describing AdS point particles of mass $m\sim\Delta$, angular momentum $\ell$ and energy $E = \Delta + 2k_{\max} + \ell$, in the semiclassical regime $\Delta\gg 1$ and $\epsilon\ll 1$. Due to the highest weight nature of the state selected by the $Y_\ell(\Omega)$ projection in \eqref{eq:amp-s}, 
the semiclassical motion is exactly on a plane determined by the angular momentum. For example, if we consider $Y_{\ell\ell}(\theta,\phi)$ in the two dimensional sphere, the uncertainty in $L_x^2+L_y^2$ is minimized. Moreover, it is much smaller than $L_z$. In the classical limit of very large angular momentum we just ignore $L_x, L_y$ and consider an orbit on the plane orthogonal to $\hat z$ that passes through the origin.
Basically, this reduces the mechanics problem to $2+1$ dimensions, so all the semiclassical physics will take place in a global $AdS_3$ geometry. In the gauge $\dot{\tau}=1$, the action is given by
\begin{equation}
  S = m\,\int d\tau\,\sqrt{\cosh^2\rho - \dot{\rho}^2 - \sinh^2\rho\,\dot{\varphi}^2}\,.
\end{equation}
Conservation of energy and angular momentum yield
\begin{equation}
\begin{aligned}
  E &= m\,\frac{\cosh^2\rho}{\sqrt{\cosh^2\rho - \dot{\rho}^2 - \sinh^2\rho\,\dot{\varphi}^2}}\,, \\
  \ell &= -m\,\frac{\sinh^2\rho\,\dot{\varphi}}{\sqrt{\cosh^2\rho - \dot{\rho}^2 - \sinh^2\rho\,\dot{\varphi}^2}}\,.
\end{aligned}
\end{equation}
Plugging the angular velocity 
\begin{equation}
  \dot{\varphi} = -\frac{\ell}{E} \frac{1}{\tanh^2\rho}\,,
\end{equation}
into the energy conservation, one can derive
\begin{equation}
  \dot{\rho}^2 = \frac{\cosh^4\rho}{\hat{E}^2} -\cosh^2\rho - \frac{\cosh^2\rho}{\tanh^2\rho} \frac{\hat{\ell}^2}{\hat{E}^2}\,,
\end{equation}
where $\hat{\ell}=\ell/m$ and $\hat{E}=E/m$. 

The insertion of the operator should correspond to the bulk particle sitting at the turning point of the orbit in the radial direction $(\dot{\rho}=0)$. Due to the presence of the angular momentum, there are two such turning points
\begin{equation}
  2\cosh^2\rho_{\pm\,\star}= \hat{E}^2-\hat{\ell}^2+1\pm\sqrt{\left(\hat{E}^2-\hat{\ell}^2+1\right)^2-4\hat{E}^2}\,.
\label{eq:rho-ell}
\end{equation}
These are the aphelion and the perihelion of the orbit. The proper matching requires to pick the aphelion, i.e. the farthest point from the origin. This satisfies our infalling intuition and agrees with the vanishing angular momentum analysis in section \ref{sec:bulk}.  

Equation \eqref{eq:rho-ell} relates the radial location of the turning point with the CFT cut-off $\epsilon$ for a given angular momentum $\ell$. In the regime $\ell\ll k_{\max}$, one recovers \eqref{eq:geom} and the bulk localization properties discussed in section \ref{sec:bulk} follow. Consider the regime $\ell \gg k_{\max}$. Since the energy satisfies
\begin{equation}
  \hat{E}^2 - \hat{\ell}^2 \sim 4\frac{k_\text{max}}{\Delta}\hat{\ell}\sim \frac{\hat \ell}{\epsilon},
\end{equation}
it follows the aphelion turning point can be approximated by
\begin{equation}
   \cosh\rho_{+\,\star} \simeq \sqrt{\frac{4 k_{max}\hat{\ell}}{\Delta}}\,.
\label{eq:aphelion}
\end{equation}
Since the fluctuations in the energy $\delta E$ at fixed $\ell$ are still controlled by $2\delta k$, we can use the CFT relative uncertainty \eqref{eq:new-rel} to derive the fluctuations in the turning point location of the bulk particle
\begin{equation}
  \sinh\rho_{+\,\star}\delta\rho_{+\,\star} \simeq \delta\left(\sqrt{\frac{4 k_{\text{max}}\hat{\ell}}{\Delta}}\right)\,.
\label{eq:comp1}
\end{equation}
In the large $\rho_\star$ limit, $\cosh\rho_\star \simeq \sinh\rho_\star$ and $\delta \rho_\star\simeq \delta k/k_{\text{max}} \simeq {\Delta}^{-1/2}$, as in \eqref{eq:compton}. Hence, despite the different location of the turning point \eqref{eq:aphelion} compared to our analysis in section \ref{sec:bulk} due to the large angular momentum $\ell$, radial fluctuations still satisfy \eqref{eq:compton}. Hence, the same conclusions regarding the bulk localization and the interpretation of these states as scattering states follow.

When $\epsilon$ is large, orbits tend to be close to circular. Indeed, the orbit is exactly circular if $\hat{E} = |\hat{\ell}|+1$, which corresponds to the condition for the descendant of the scalar field to have the minimum energy at angular momentum $\ell$, i.e. $E=\ell+\Delta$. For this value, one gets $\cosh \rho_\star=\sqrt{\hat{E}}$. Orbits are approximately circular if  $2 k_{\text{max}}/m$ is very small compared to $\hat \ell$. This usually occurs if $\epsilon$ is large.

One might also be interested in placing a local operator in a small region that produces an initial configuration at a specific angle, with a corresponding angular momentum $\ell$, so that instead of a smeared particle in the angular directions as would be appropriate for a specific scattering state in some partial wave, one gets instead a good classical initial geometric condition for some particular geodesic. This can be done by using the line integral
\begin{equation}
\CO_{\epsilon,\ell }(\theta, t) \simeq \int d\theta' \exp(i \ell \theta') f(\theta-\theta') \CO_{\Delta, \epsilon}(\theta',t)
\end{equation}
where $f(\theta-\theta')$ is a function of compact support on a small region and approximately Gaussian with an $\epsilon$ dependent width $\delta \theta$ so that $\delta \theta \ell \gg 1$, and the line element is aligned in the classical direction that we want to specify for the particle velocity that is represented by the angular momentum.
This way the integral will produce an uncertainty in the angular momentum that is small, $\delta \ell/\ell \ll1$, as well as a small uncertainty in the initial condition $\delta \theta$.
Similarly to the radial geodesics for local operator insertions, one would also produce a small uncertainty in the transverse angles to the angular momentum, as derived from the two point function \eqref{eq:trans_unc}.

\section{Conclusions}
\label{sec:conc}

In this work we have constructed regularized CFT scalar primary operators 
\begin{equation}
  \CO_{\Delta,\epsilon}(\theta,t)= \exp(-\epsilon H )\, \CO_{\Delta}(\theta,t)\exp(+\epsilon H )
\label{eq:regulator}
\end{equation}
whose insertions acting on the CFT vacuum generate the primary state and its generic descendants through the operator-state correspondence. The CFT regulator $\epsilon$ makes the sum of the amplitudes of these states finite, so that the state
$\CO_{\Delta,\epsilon}(\theta,t)|0\rangle$ is normalizable. 

Working in the limit of large conformal dimension $(\Delta\gg 1)$,  these amplitudes are well approximated by gaussian distributions with sharp widths. We have shown, in this same limit, that these states correspond to semiclassical bulk particles sitting at rest at the turning point of the relevant geodesic associated with the given state. This perspective provides a geometrical interpretation for the CFT regulator $\epsilon$ in terms of the turning point location of the geodesic. CFT gaussian fluctuations controlling the amplitude distribution of the state are then mapped into bulk localization properties. We find these bulk particles are well localized within the AdS radius and the Compton wave-length $\lambda_{\text{c}}$ of the particle for generic descendants (see \eqref{eq:compton} and the discussion below \eqref{eq:comp1}). We have argued that projecting operators \eqref{eq:regulator} with different kernels
\begin{equation}
 \int d\Omega\, \CO_{\Delta, \epsilon}(\theta, t)\ket 0\,, \quad \int d\Omega\, Y_\ell(\Omega) \CO_{\Delta, \epsilon}(\Omega, t)\ket 0\,, \quad \int d\theta' \exp(i \ell \theta') f(\theta-\theta') \CO_{\Delta, \epsilon}(\theta',t) \ket 0
\end{equation}
corresponds to choosing different bulk geodesics accounting for the different symmetries of the states and their localization properties.

Furthermore, for descendants carrying no angular momentum, we have also shown their amplitudes are compatible with an euclidean bulk tunneling interpretation from the boundary to the turning point that also accounts for the CFT euclidean time evolution appearing in the regulator \eqref{eq:regulator}. It is natural to assume that the same will be true in the presence of angular momentum and we are currently trying to understand this 
issue.

There are some natural extensions of our work. First, one could consider the insertion of similarly regularized operators in less symmetric CFT states. These could either correspond to smooth holographic gravity duals or to black holes. In the latter case, the insertion of such regularized operators has been considered in the context of holographic quenches and the evolution of entanglement and scrambling at finite temperature, for example in the works \cite{Caputa:2014eta,Caputa:2015waa,Stikonas:2018ane,localq}.  It would be particularly interesting to study the localisation properties of the holographic bulk particle at finite temperature in the large angular momentum regime since one expects a large effective potential arising in this case.

Second, the interpretation of our insertions as scattering states raises the possibility of using their holographic description to provide potentially interesting relations to the bootstrap programme. In principle, we could study the scattering from a heavy object at different impact parameters holographically by the study of geodesics on the relevant gravity duals, where we could attempt to learn about anomalous dimensions of double trace operators and complement the work \cite{Aprile:2017bgs,Alday:2017xua} for operators of higher dimension.

Another natural question to ask is how localized the operators $\CO_{\epsilon}(\theta,t)$ are 
on the boundary.  After all, they are non-local. The natural place to ask this question is in terms of entanglement wedge reconstruction. We can ask what is the smallest region of the boundary at fixed $t$
where such an operator can be localized within the local algebra of operators of the region.
The natural answer should be that if the particle is created inside the entanglement wedge of 
that region, then a close approximation to the operator should be possible in the corresponding boundary \cite{Dong:2016eik}. Because global AdS is static and stationary, this requires solving for 
Ryu-Takayanagi surfaces \cite{Ryu:2006bv} at fixed $t$. For disk shaped regions centered at $\theta'$, these will look like a spherical dome (see for example the arguments in 
\cite{Berenstein:1998ij}). 
The minimal disk shaped region will occur, by symmetry,  when the disk is centered at $\theta'=\theta$, and whenever the corresponding $\rho_*$ is barely inside the dome. For small $\epsilon$ we can ignore global $AdS$
and instead work in the local Poincare coordinates on the boundary. The rotational symmetry of the Euclidean  coordinates between the $\tau$ direction and the local $\delta \theta$ direction show that the corresponding region is of size $\delta\theta\simeq \epsilon$, as the domes are foliated by geodesics in AdS.

\acknowledgements
D.B. thanks  S. Giddings and A. Miller for many discussions over the years.
D.B. is very grateful to the Visiting Programme at the Higgs Center for Theoretical Physics in Edinburgh for partial funding support and for its hospitality. The work of D.B. is supported in part by the Department of Energy under grant {DE-SC} 0011702.

\end{document}